\documentclass{article}
\usepackage{spconf,amsmath,graphicx}

\newlength{\bibitemsep}
\newlength{\bibparskip}
\let\oldthebibliography\thebibliography
\renewcommand\thebibliography[1]{%
	\oldthebibliography{#1}%
	\setlength{\parskip}{\bibitemsep}%
	\setlength{\itemsep}{\bibparskip}%
}

\usepackage{fancyhdr}
\title{QUANTIZED DECODER IN LEARNED IMAGE COMPRESSION FOR DETERMINISTIC RECONSTRUCTION}
\name{Esin Koyuncu\textsuperscript{1,2}, Timofey Solovyev\textsuperscript{2}, Johannes Sauer\textsuperscript{2}, Elena Alshina\textsuperscript{2}, Andr\'{e} Kaup\textsuperscript{1}}
\address {{\textsuperscript{1}Friedrich-Alexander-Universit\"{a}t Erlangen-N\"{u}rnberg, Erlangen, Germany}  \\
	{\textsuperscript{2}Huawei Technologies, Munich, Germany} }

\usepackage{xcolor, color, soul}
\sethlcolor{yellow}
\usepackage{multirow}
\usepackage{url}
\usepackage{booktabs}
\usepackage{hyperref} 

\fancypagestyle{FirstPage}{
	\fancyhf{}
	\lhead{\scriptsize{\copyright 2024 IEEE. Personal use of this material is permitted. Permission from IEEE must be obtained for all other uses, in any current or future media, including reprinting/republishing this material for advertising or promotional purposes, creating new collective works, for resale or redistribution to servers or lists, or reuse of any copyrighted component of this work in other works.} 
	}
}
\begin{document}
%
\maketitle
\begin{abstract}
Learned image compression has a problem of non-bit-exact reconstruction due to different calculations of floating point arithmetic on different devices. This paper shows a method to achieve a deterministic reconstructed image by quantizing only the decoder of the learned image compression model. From the implementation perspective of an image codec, it is beneficial to have the results reproducible when decoded on different devices. In this paper, we study quantization of weights and activations without overflow of accumulator in all decoder subnetworks. We show that the results are bit-exact at the output, and the resulting BD-rate loss of quantization of decoder is 0.5\% in the case of 16-bit weights and 16-bit activations, and 7.9\% in the case of 8-bit weights and 16-bit activations. 
\end{abstract}
\section{INTRODUCTION}
\label{sec:intro}

The demand for storage and transmission of multimedia data rapidly increases, therefore continuous improvement of image/video compression algorithms and their deployment is desired. JPEG \cite{wallace1991jpeg}, JPEG2000 \cite{skodras2001jpeg}, HEVC {\cite{sullivan2012overview} and the state-of-the-art VVC \cite{bross2021overview} are widely used standards based on classical compression methods. End-to-end learning based methods \cite{Ball2016EndtoendOI, Ball2018VariationalIC, Minnen2018JointAA, cheng2020learned, liu2023learned} have been studied extensively, some outperforming VVC. It is not a straightforward task to deploy the learned image codecs on devices due to the non-deterministic behaviour of neural networks and complexity of the computations. 

Neural network based image codecs are trained with floating point (fp32) arithmetic. Storing the resulting models requires a much larger memory than is needed for storing integer models (int8/int16) with the same number of parameters. Also the run time of floating point operations is higher than the run time of integer operations. Quantization of the neural network as a solution to the complexity problem has been studied by previous work in  \cite{hong_efficient_2021, sun2022-Q-LIC, le_mobilecodec_2022, sun_end--end_2020, sun_learned_2021pcs, balle2018integer}. Another problem of learned image compression are the totally corrupted images occurring at the cross platform tests. Quantization of the weights and activations of the entropy part to fix this issue have been proposed by \cite{sun_learned_2021pcs, balle2018integer, koyuncu2022device}.

Our aim is to achieve bit-exact reconstructed images at decoder when running on different platforms. For this it is sufficient to only quantize the neural network components at the decoder. For many of the use cases of image coding, the decoder of the image codec is fixed, while the encoder design is flexible. Therefore, we quantize only the decoder in order to provide bit exact reconstruction and keep the encoder side free for complex hardware implementation. We reuse the idea in \cite{koyuncu2022device} for post-training quantization of weights and activations without overflow and extend it to all decoder subnetworks of the base operation point of JPEG AI VM (verification model) \cite{VM_software} as shown in Fig. \ref{fig:jpegai_architecture}. 
We show that by quantizing only decoder networks by reusing the method to eliminate the overflow of the 32-bit accumulator for 16-bit weights and 16-bit activations, we could achieve the deterministic decoded reconstructed images with a minor loss. Also, we provide more experiments to study the 8-bit weights with 16-bit activations, and also to compare the effect of choosing the optimal clipping range instead of clipping to the maximal range for both cases with 8-bit and 16-bit weights. In Section \ref{sec:related work}, we summarize the previous related work and compare it to ours. The overall description of the learned image coding architecture is given in Section \ref{ss:codec}, and its quantized components are explained in Section \ref{ss:quantized_part_of_codec}. In Section \ref{ss:quantization_details}, the method of quantization is described. The experimental results are provided in Section \ref{sec:Experiments} and conclusions are given in Section \ref{sec:discussion}.
\begin{figure}[hbtp]
	\centerline{\includegraphics[width=0.9\columnwidth]{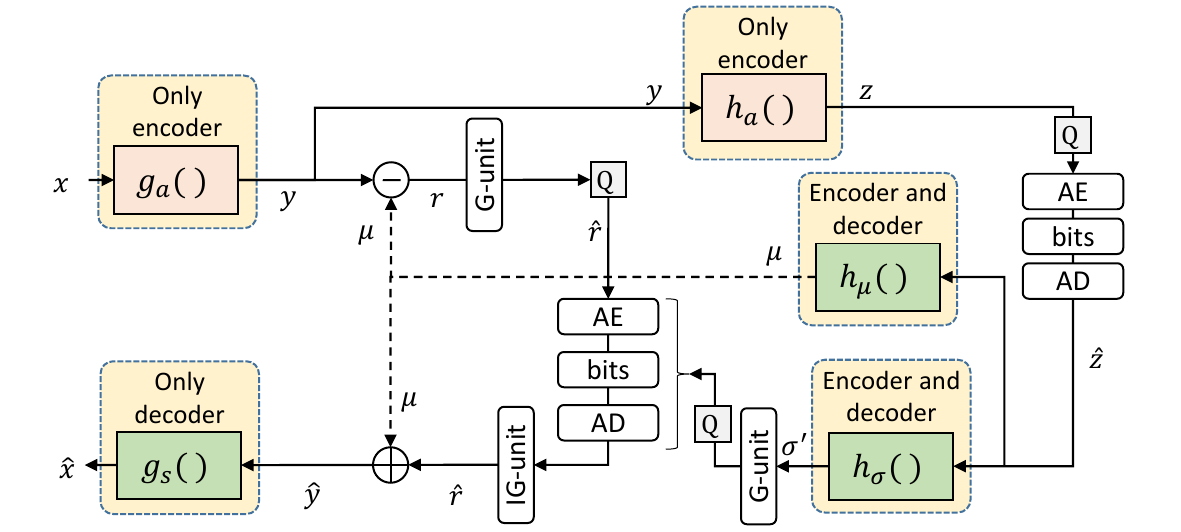}}
	\caption{High-level architecture diagram of the JPEG AI base operating point. 
		Encoder side subnetworks are $g_a$, $h_a$, $h_{\mu}$, $h_{\sigma}$. Decoder side subnetworks are $g_s$, $h_{\mu}$, $h_{\sigma}$.}	
	\label{fig:jpegai_architecture}
\end{figure}

\section{RELATED WORK} \label{sec:related work} 

In \cite{sun_learned_2021pcs, balle2018integer, koyuncu2022device} we see the quantization of weights and activations in entropy part as a solution to the visible cross platform decoding error. 
In this paper we extend the post training quantization method of 16-bit weights and activations with no-overflow of 32-bit accumulator in \cite{koyuncu2022device}. While \cite{koyuncu2022device} quantizes the entropy part only, this work applies quantization to the whole decoder. We target a zero deviation and achieve exact reconstructed images at cross platform tests while keeping the performance loss compared to the floating point model minimal. Although the quantization of only the entropy part as in \cite{koyuncu2022device} eliminates the broken images error at the decoder, there still exists a tiny deviation of the quality of the reconstructed images due to the remaining unquantized layers of the decoder, which is not visible visually but can be measured numerically. This is not a critical problem for the image compression task, but for some error-critical applications such as medical image processing, or the video compression task, where the reconstructed image typically becomes a reference for the prediction of the other frames, it is desired to have no deviation. Also we show the results of quantizing only the hyper decoder networks. This allows exact decoding of $\hat{y}$ and can be interesting for computer vision tasks which can use decoded latent tensors $\hat{y}$ directly as input, but do not need the reconstructed image.

In \cite{hong_efficient_2021, sun2022-Q-LIC, le_mobilecodec_2022, balle2018integer}, both weights and activations of the whole codec are quantized. In \cite{sun_end--end_2020} only weights (not activations) of the whole codec are quantized. In \cite{sun_learned_2021pcs} weights of the whole codec and activations of only the entropy part are quantized.
We quantize weights and activations of whole decoder, i.e., decoder layers in main and entropy part. It is not necessary to quantize all layers in the encoder to achieve the deterministic reconstructed image. This way we allow more complexity at the calculations but the problem of the bit-exact reconstruction is still simple to solve by a post-training quantization method. In common use cases such as streaming, encoding is done once on a device, and decoding is done on many devices. With proposed decoder quantization, for the same input image, encoders on different devices can generate bitstreams with small variation. However, we ensure that given the same bitstream, decoders on different devices generate bit-exact reconstructed images.

\section{PROPOSED METHOD}
\subsection{Learned image codec} \label{ss:codec}
\begin{figure*}[t]
	\centerline{\includegraphics[width=15cm]{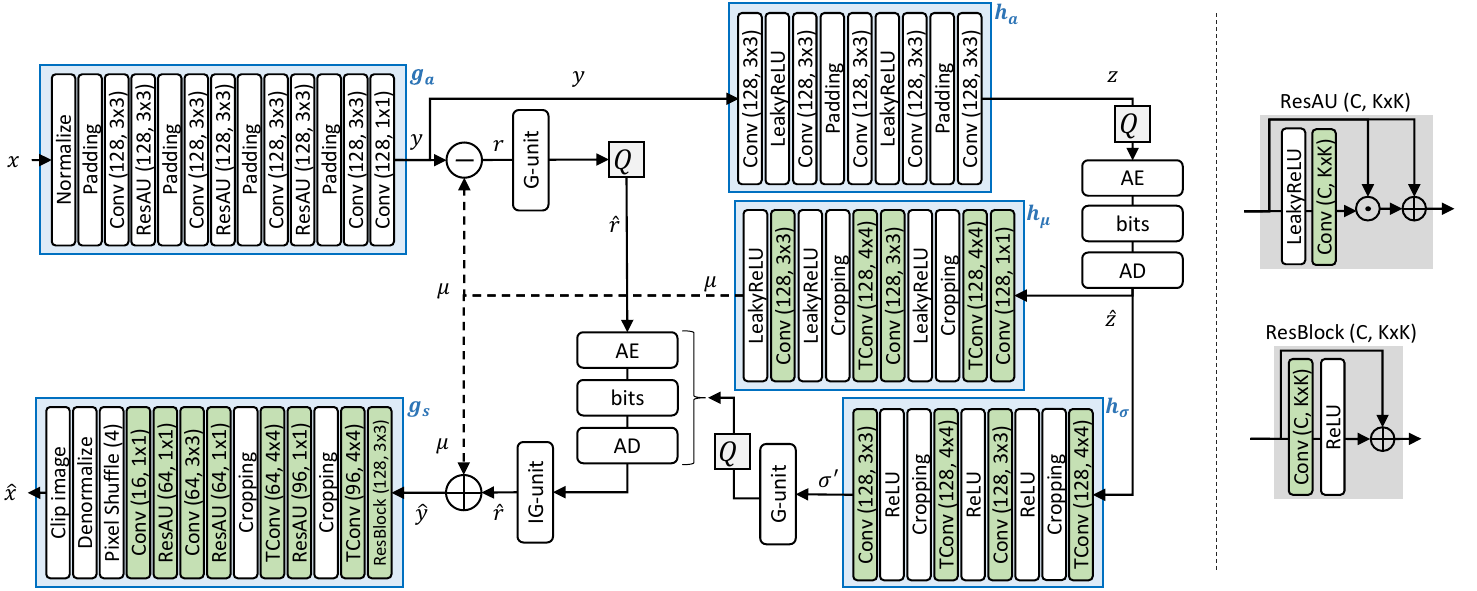}}
	\caption{A detailed architecture diagram of JPEG AI base operating point. Quantized layers are marked with green colour.}
	\label{fig:overall_detailed}
\end{figure*}
As a baseline image codec, we use the base operating point of JPEG AI VM software version 3.4.0 \cite{VM_software}. The JPEG AI VM has two different operating points, high and base, with a different trade-off between complexity and performance. The base operating point architecture consists of an entropy model without context and it has simplified layers and activation functions compared to the high operating point. The architecture of the base operating point for one colour component is shown in Fig. \ref{fig:jpegai_architecture}. This codec processes colour components separately. Each subnetwork consists of two separate components of subnetworks, one processing the luma channel (Y) and one processing the two chroma channels (U, V). The decoder consists of three main subnetworks, a hyper scale decoder, $h_{\sigma}$, a hyper decoder, $h_{\mu}$, and a main decoder, $g_s$; the subnetworks are highlighted in green in Fig. \ref{fig:jpegai_architecture}.
There are four models which are trained with different rate-distortion (RD) loss and overall they produce five different quality points during inference. G-unit and IG-unit are the trained vectors for gain and inverse gain unit, which provide the variable rate functionality as in \cite{gvae_cui2021asymmetric} and thus derive the highest (fifth) rate point from the model of the fourth rate point. The JPEG AI VM software has additional tools to improve the compression quality but all these additional tools are turned off during our implementation, so the coding pipeline is reduced to the neural network layers which are shown in Fig. \ref{fig:overall_detailed}. The explicit block diagram of the light-weighted residual activation unit (ResAU) and the light-weighted residual block (ResBlock) are shown on the right-hand-side of Fig. \ref{fig:overall_detailed}.

\subsection{Quantized part of the codec} \label{ss:quantized_part_of_codec}
We use the floating point trained models as the anchor, and apply the quantization to the decoder subnetworks. The layers which are highlighted in green in Fig. \ref{fig:overall_detailed}  are the convolutional layers to be quantized. The $h_{\sigma}$ is the subnetwork which is calculating the entropy parameters and is shared between the encoder and decoder. From the previous work \cite{balle2018integer, sun_learned_2021pcs, koyuncu2022device}, we know that it is necessary to quantize the entropy part of
the subnetwork  in order to eliminate
catastrophic decoded images. In this paper, we extend the partial quantization of the codec to all the elements of the decoder, with the aim of deterministic calculation of the reconstructed image $\hat{x}$ at the encoder and decoder side. We replace the LeakyReLU activations in the hyper decoder $h_{\mu}$ with an alternative implementation using bit shift. The subnetworks which only exist at the encoder are the main encoder $g_a$ and the hyper encoder $h_a$; these subnetworks are kept as floating point in our study. 

\subsection{Methodology for quantization of decoder subnetworks weights and activations}
\label{ss:quantization_details}

Our work is based on the post training quantization method in \cite{koyuncu2022device}.  This method chooses the optimal per-channel quantization parameters for weights and per-layer quantization parameters for activations while it takes care to avoid an overflow of the accumulator register (guaranteed theoretically for any potential input). The input clipping range and accumulator size are used to calculate the maximal safe precision of the weights per channel.

We quantize the decoder subnetworks sequentially in three steps. The first step is the quantization of the entropy subnetwork, $h_{\sigma}$, the second step is the quantization of the $h_{\mu}$, and the third step is the quantization of the main synthesis transform $g_s$. When going to the next step, we use the model with the quantized networks of the previous step.

We always quantize the activations to 16-bit. The accumulator size is 32-bit, and whenever there is a bias term it is quantized to 32-bit. We show two sets of results, one with 8-bit quantized weights and one with 16-bit quantized weights. We quantize the $h_{\sigma}$ weights (step one) to 8-bit only since its loss is already minor. For simplicity, in this paper we say that all decoder subnetworks are quantized with 16-bit weights, even when the $h_\sigma$ weights are 8-bit. Also for both cases, we study the effect of the  different activation clipping ranges.

In \cite{koyuncu2022device} weights and activations are both 16-bit. There is one difference of the weights quantization method from \cite{koyuncu2022device} if we apply 8-bit quantization to weights. Instead of directly scaling the weights with the calculated factor that eliminates the overflow $2^{-k_j}$ as in \cite{koyuncu2022device}, we use this as an upper bound and search the power of two of scaling factor in interval $[0, k_j]$ that results in the minimal quantization error for the weights. We observe that this is necessary to minimize the clipping error. Because we clip the 8-bit weights to the interval $[-2^7, 2^7-1]$ instead of $[-2^{15}, 2^{15}-1]$ as we do for the 16-bit weights case, scaling with the maximum available scaling factor is not ideal for the smaller clipping range.

Activations precision and clipping range are chosen with a search which optimizes the average BD-rate \cite{bjontegaard2001calculation} compared to the anchor with unquantized layers. We search for the optimal parameters layer by layer. The decision is made based on the average BD-rate of a calibration data set calculated by two distortion metrics as in \eqref{eqn:mixed_bd_Rate}, where combined YUV-PSNR has weights of 0.8 for Y, and 0.1 for U and V components.

\begin{equation}\label{eqn:mixed_bd_Rate}
	\resizebox{0.9\columnwidth}{!} {
$	\mathrm{BD}  \text{-} \mathrm{rate} =\frac { \mathrm{BD}  \text{-} \mathrm{rate(YUV} \text{-} \mathrm{PSNR)} + \mathrm{BD}  \text{-} \mathrm{rate(MS} \text{-} \mathrm{SSIM)}} {2} $
}
\end{equation} 

Since the target is to achieve bit-exact reconstructed images at the cross platform tests, we define an error measure to express the bit-exactness in a quantitative manner. We run the encoder on one device and the decoder on another device and measure the mean square error (MSE) between the reconstructed images $\hat{x}$ at encoder and decoder side. The way that we calculate cross-platform deviation is 

\begin{equation}
	\label{eqn:rec_dec_error}
	\mathrm{MSE_{enc,dec}} = \sum_{c \in \{Y, U, V\}} \mathrm{MSE} (\hat{x}_{c, \mathit{enc}}, \hat{x}_{c, \mathit{dec}})
\end{equation}

where $\hat{x}_{c, \mathit{enc}}$ denotes the reconstructed image calculated at the encoder, and $\hat{x}_{c, \mathit{dec}}$ denotes the reconstructed image calculated at the decoder for colour component $c$.

\section{EXPERIMENTAL RESULTS}
\label{sec:Experiments}

\begin{table*}[]
	\centering
	\caption{BD-rate (\%) results of quantized sub-parts of decoder compared to the anchor of floating point model}
	\vspace*{2mm}
	\label{tab:bd-rate-results}
	\resizebox{0.81\width}{!}{%
		\begin{tabular}{@{}lllccccc@{}}
			\toprule
			\multicolumn{3}{c}{\textbf{Quantized weights bit depth }} &    \multicolumn{5}{c}{\textbf{BD-rate calculated based on the distortion metric}} 
			\\ \cmidrule(lr){1-3}\cmidrule(lr){4-8}
			{\boldmath{$h_{\sigma}$}} & 			{\boldmath{$h_{\mu}$}} & 			{\boldmath{$g_s$}} &			
			{\small{\textbf{MS-SSIM}}} &
			{\small{\textbf{Y-PSNR}}} &
			{\small{\textbf{U-PSNR}}} &
			{\small{\textbf{V-PSNR}}} &
			{\small{\textbf{YUV-PSNR}} } \\ \toprule
			{int8} &
			{-} &
			{-} &
			{0.00} &
			{0.02} &
			{0.62} &
			{1.56} &
			{0.23} \\ \midrule
			{int8} &
			{int16} &
			{-} &
			{0.01} &
			{0.06} &
			{0.50} &
			{1.45} &
			{0.24}
			\\ \midrule
			{int8} &
			{int16} &
			int16 &
			{0.46} &
			{0.57} &
			{0.96} &
			{2.30} &
			{0.78}
			\\ \midrule
			{int8} &
			{int8} &
			{-} &
			{0.85} &
			{0.96} &
			{1.76} &
			{3.06} &
			{1.25} \\ \midrule
			{int8} &
			{int8} &
			{int8} &
			{7.86} &
			{7.08} &
			{5.04} &
			{8.14} &
			{6.98} \\ \bottomrule
		\end{tabular}%
}
	\end{table*}
 The trained models of the JPEG AI VM base operation point are the anchor of the experiments. Then, we apply quantization to the weights and activations of the subnetworks of these models. We use the training, validation and test data which are provided by the JPEG AI common test conditions \cite{JPEGAI_datasets, jpegaiCTC}. For quantization we use the calibration set as in \cite{VM_software}. It is a set of 16 images of various resolutions and content picked from validation set. For the results evaluation, we run the inference on the 50-image JPEG AI test set. We measure the average BD-rate deviation from the floating point model as well as the deviation between reconstructed images of  
 encoder and decoder. We show how much each decoder network component contributes to the reconstructed image deviation. 

In Table \ref{tab:bd-rate-results}, we present the results of the BD-rate loss of quantization. If we quantize all decoder subnetworks with 16-bit weights and 16-bit activations, the performance loss is 0.46\% and 0.78\% of BD-rate based on MS-SSIM and YUV-PSNR, respectively. If we quantize all decoder subnetworks with 8-bit weights and 16-bit activations, the performance loss is 7.86\% and 6.98\% of BD-rate based on MS-SSIM and YUV-PSNR, respectively.
 We also see the loss of the intermediate steps (`-' means subnetwork is not quantized) and observe that the larger loss is introduced by the synthesis transform quantization.

We show the reconstructed image error of cross platform encoding and decoding in Table \ref{tab:xplatform_results}. If the model is not quantized at all, there is a huge deviation at the reconstructed tensors which are calculated at the encoder and decoder side. If we quantize only the entropy part, $h_{\sigma}$, we can decrease this error significantly but it is still not zero. If we quantize additionally the hyper decoder  $h_{\sigma}$, the deviation decreases by a small amount. Only if we quantize all three decoder networks, we can achieve zero error, and confirm that the reconstructions are calculated exactly the same.
\begin{table}[tb]
	\centering
	\caption{MSE between reconstructed picture at encoder and decoder at cross platform tests \\ GPU: NVIDIA GeForce RTX 4090 \\ CPU: Intel(R) Core(TM) i9-10980XE CPU}
	\vspace{3mm}
	\label{tab:xplatform_results}
	\resizebox{0.8\width}{!}{
		\begin{tabular}{@{}lllcc@{}}
			\toprule
			\multicolumn{3}{c}{\textbf{Quant. subnetwork}} & \multicolumn{2}{c}{\textbf{Tested device}} 
			\\ \cmidrule(lr){1-3}\cmidrule(lr){4-5}
			{\boldmath{$h_{\sigma}$}} & 			{\boldmath{$h_{\mu}$}} & 			{\boldmath{$g_s$}} &			\textbf{\begin{tabular}[c]{@{}c@{}}enc: GPU  dec: CPU\end{tabular}} & \textbf{\begin{tabular}[c]{@{}c@{}}enc: CPU dec: GPU\end{tabular}} \\ \toprule
			
			{-} &
			{-} &
			{-} &
			{2.08E+04} &
			{2.10E+04}
			\\ \midrule
			{int8} &
			{-} &
			{-} &
			{2.15E-09} &
			{2.15E-09}
			
			\\ \midrule
			{int8} &
			{int16} &
			{-} &
			{1.02E-09} &
			{1.02E-09}
			
			\\ \midrule
			{int8} &
			{int8} &
			{-} &
			{1.04E-09} &
			{1.04E-09}
			
			\\ \midrule
			{int8} &
			{int16} &
			{int16} &
			{\textbf{0}}&
			{\textbf{0}}
			\\ \midrule
			
			{int8} &
			{int8} &
			{int8} &
			{\textbf{0}}&
			{\textbf{0}}\\ \bottomrule
		\end{tabular}%
	}
\end{table}

Additionally, we study the effect of different activation clipping ranges compared to a fixed clipping range. We test two variants, one with 8-bit weights and one with 16-bit weights. Comparing the results shown in Table \ref{tab:fixed-clip-results} with the flexible clipping results in Table \ref{tab:bd-rate-results}, we observe that clipping to different values can help the performance significantly when quantizing with 16-bit weights and activations, while it does not help much with 8-bit weights and 16-bit activations case. This is because the risk of overflow of the 32-bit accumulator is much higher when the weights and input tensors both have a bit depth of 16 bits.

\begin{table}[htb]
	\setlength{\tabcolsep}{0.33em}
	\centering
	\caption{BD-rate (\%) results compared to the of floating point model anchor when activations are clipped to fixed range}
	\vspace{3mm}
	\label{tab:fixed-clip-results}
	\resizebox{\columnwidth}{!}{
		\begin{tabular}{@{}lllccccc@{}}
			\toprule
			\multicolumn{3}{c}{\textbf{Quant. subnetwork}} &    \multicolumn{5}{c}{\textbf{BD-rate calc. based on the distortion metric}} 
			\\ \cmidrule(lr){1-3}\cmidrule(lr){4-8}
			{\boldmath{$h_{\sigma}$}} & 			{\boldmath{$h_{\mu}$}} & 			{\boldmath{$g_s$}} &			
			\small{\textbf{MS-SSIM}} &
			\small{\textbf{Y-PSNR}} &
			\small{\textbf{U-PSNR}} &
			\small{\textbf{V-PSNR}} &
			\small{\textbf{YUV-PSNR}} \\ \toprule

			{int8} &
			{int16} &
			int16 &
			{2.56} &
			{1.51} &
			{1.87} &
			{2.95} &
			{1.69}
			\\ \midrule
			
			{int8} &
			{int8} &
			{int8} &
			{7.85} &
			{7.07} &
			{5.07} &
			{8.17} &
			{6.98} \\ \bottomrule
		\end{tabular}%
	}
\end{table}
\section{DISCUSSION}
\label{sec:discussion}
In this paper, with the aim to achieve deterministic output pictures at encoder and decoder at cross platform tests, we present a method which quantizes the decoder subnetworks of a learned image codec JPEG AI base operating point. We study a simple post training quantization with per layer activation quantization and per channel weights quantization which eliminates the overflow. With this approach, we can keep the original floating point parameters of the encoder-only subnetworks of the codec, thereby not limit encoder RD optimization algorithms. 

In the case when all decoder subnetworks are quantized using 16-bit weights and activations, we show that the loss from the floating point model is small and we can achieve bit exact reconstructed images at the cross platform tests. In the case of 8-bit weights and 16-bit activations, we still show the benefit of deterministic calculation of reconstructed image but the loss relative to the floating point models is higher. These two solutions at two different settings can be interesting for different purposes. For research tasks such as learning based or hybrid video compression, it might be beneficial to stop the deviation of the decoded reconstructed picture while keeping the performance as close as possible to floating point trained model since it can eliminate the error accumulation to the predicted frame. On the other hand, for tasks such as optimizing the complexity of the decoder there might be an interest to use 8-bit weights since it decreases the storage load. The speed benefit of our work is limited when activations are quantized to 16-bit since GPU acceleration of tensor operations is only available for 8-bit, 4-bit or binary input operands \cite{nvidia_whitepaper} today. However, smartphone NPUs support the inference with 8-bit weights and 16-bit activations \cite{snapdragon_npu}.
\vfill\pagebreak

\bibliographystyle{IEEEbib}
\bibliography{strings,refs}

\end{document}